\documentclass[aps,prd,twocolumn,notitlepage,
nofootinbib,nobibnotes,superscriptaddress,preprintnumbers]{revtex4-1}

\usepackage{graphicx,xcolor}
\usepackage{physics}
\usepackage{amsmath,amssymb}
\usepackage{enumitem}   
\usepackage{dcolumn}
\usepackage[hidelinks]{hyperref}
\usepackage[normalem]{ulem}
\usepackage{dblfloatfix}

\newcommand{\avg}[1]{\left\langle#1\right\rangle}

\newcommand{\Adyn}{\mathcal{A}^\mathrm{dyn}}

\begin{document}
\title{Effective spectral function of vector mesons via lifetime analysis}
\author{Renan Hirayama}
\affiliation{Helmholtz Forschungsakademie Hessen für FAIR (HFHF), GSI Helmholtzzentrum für Schwerionenforschung, Campus Frankfurt, Max-von-Laue-Str. 12, 60438 Frankfurt am Main, Germany
}%
\affiliation{Frankfurt Institute for Advanced Studies, Ruth-Moufang-Strasse 1, 60438 Frankfurt am Main, Germany}

\author{Jan Staudenmaier}\affiliation{Frankfurt Institute for Advanced Studies, Ruth-Moufang-Strasse 1, 60438 Frankfurt am Main, Germany}%
\affiliation{Institut f\"{u}r Theoretische Physik, Goethe Universit\"{a}t, Max-von-Laue-Strasse 1, 60438 Frankfurt am Main, Germany}
\affiliation{Department of Physics and Astronomy, Wayne State University, Detroit MI 48201, United States}

\author{Hannah Elfner}%

\affiliation{Helmholtz Forschungsakademie Hessen für FAIR (HFHF), GSI Helmholtzzentrum für Schwerionenforschung, Campus Frankfurt, Max-von-Laue-Str. 12, 60438 Frankfurt am Main, Germany
}%
\affiliation{Frankfurt Institute for Advanced Studies, Ruth-Moufang-Strasse 1, 60438 Frankfurt am Main, Germany}
\affiliation{Institut f\"{u}r Theoretische Physik, Goethe Universit\"{a}t, Max-von-Laue-Strasse 1, 60438 Frankfurt am Main, Germany}
\affiliation{GSI Helmholtzzentrum f\"{u}r Schwerionenforschung, Planckstr. 1, 64291 Darmstadt, Germany}

\date{\today}

\begin{abstract}

Effective spectral functions of the $\rho$ meson are reconstructed by considering the lifetimes inside different media using the hadronic transport SMASH (Simulating Many Accelerated Strongly-interacting Hadrons). Due to inelastic scatterings, resonance lifetimes are dynamically shortened (\emph{collisional broadening}), even though the employed approach assumes vacuum resonance properties. Analyzing the $\rho$ meson lifetimes allows to quantify an effective broadening of the decay width and spectral function, which is important in order to distinguish dynamical effects from additional genuine medium modifications to the spectral functions, indicating e.g. an onset of chiral symmetry restoration.
The broadening of the spectral function in a thermalized system is shown to be consistent with other theoretical calculations. The effective $\rho$ meson spectral function is also presented for the dynamical evolution of heavy-ion collisions, finding a clear correlation of the broadening to system size, which is explained by an observed dependence of the width on the local hadron density. Furthermore, the difference in the results between the thermal system and full collision dynamics is explored, which may point to non-equilibrium effects. 
\end{abstract}

\maketitle

\section{Introduction}

Under extreme thermodynamic conditions the restoration of chiral symmetry is expected in Quantum Chromodynamics (QCD). The spectral functions of chiral partners become degenerate as a consequence. The most  accessible of these chiral pairs is composed of $(\rho,a_1)$, the vector and axial-vector mesons with the lowest masses. In particular, the $\rho$ meson's direct dilepton decay allows direct experimental access to its in-medium modifications of hadronic properties. Therefore, a large experimental and theoretical effort has been undertaken to study the behavior of the spectral function of the $\rho$ meson, in and out of thermal equilibrium.

Experimentally, QCD and particularly the properties of resonances in a hot and dense medium are studied through nucleus-nucleus collisions. Dileptons are particularly attractive to study in-medium properties, since they are not subject to the strong force and only interact electromagnetically with the surrounding fireball, escaping the medium essentially undisturbed. 

The dilepton emission was therefore studied by various experiments, starting with the CERES/NA45 experiment at CERN-SPS at high beam energies  \cite{wessels2003latest}, exhibiting an excess attributed to the direct radiation of the fireball. Afterwards, the NA60 experiment \cite{arnaldi2006first} settled the debate about the nature of this excess: it was found consistent with a strong broadening of the $\rho$ spectral function with no apparent mass shift. The latter being predicated by a handful of theoretical models as the `dropping mass' scenario along with an increasing width \cite{pisarski1982phenomenology,hatsuda1992qcd,li1995enhancement,leupold1998qcd}.  
More recently, the HADES experiment at GSI investigated the kinematic regime of low beam energies for p+p, p+A, and A+A collisions. An excess in the measured electron pair yield $M_{ee}\sim 0.15$-$0.5$ GeV/c$^2$ \cite{agakichiev2007dielectron,agakishiev2012first,agakishiev2012inclusive, agakishiev2012inclusive2,agakishiev2011dielectron}, in agreement with previous results from the DLS experiment at BEVALAC \cite{porter1997dielectron} was found, again showing an excess yield connected to strong coupling of the $\rho$ meson to the baryonic sector \cite{agakichiev2010origin, rapp2009chiral}.
Future experiments, such as CBM at FAIR, will add more high-quality experimental data on dilepton emission in the high net-baryon density region, by probing the intermediate energy range \cite{hohne2014measurement}. 

Different theoretical techniques are employed to study the vector meson spectral functions. A well-established approach ensues from hadronic many-body theory \cite{rapp1997rho,rapp1999low,rapp2005vector,peters1998spectral}, evaluating the $\rho$ propagator including several modifications to its self-energy. Convolving it with a simple fireball model resulted in a dilepton yield consistent to SPS data \cite{van2008dilepton,van2006comprehensive}. One more recent advance comes from the Functional Renormalization Group approach, a non-perturbative framework capable of taking critical fluctuations into account \cite{tripolt2014spectral,rennecke2015vacuum}, and therefore is suited to study medium modifications in cold and dense nuclear matter close to the liquid-gas phase transition, as well as the effects of the corresponding critical endpoint \cite{jung2017medium,tripolt2021vector}. 

Transport approaches allow to connect theoretical approaches for the spectral function to experimental measurements. They describe the full evolution in a collision and, therefore, have the advantage of allowing direct access to the spectral information of particles at all times. In some approaches, the resonances can propagate \emph{off-shell}, and their spectral function can dynamically change, with the constraint of a vacuum behavior in the absence of surrounding matter. One example is (P)HSD \cite{cassing2000semiclassical,bratkovskaya2008dilepton}, which computes the spectral function at each time step via transition amplitudes; another is GiBUU \cite{buss2012transport,larionov2020dilepton}, in which the invariant mass of the particle becomes an independent variable, determined self-consistently through the evolution. Other transport approaches, such as UrQMD \cite{bass1998microscopic}, deal with \emph{on-shell} resonances always assuming vacuum properties. SMASH (Simulating Many Accelerated Strongly-interacting Hadrons) \cite{weil2016particle}, the hadronic transport used for this work, falls in this category as well.

A possibility to supplement direct medium-modifications to the spectral function to vacuum hadronic transport approaches is the coarse-graining method, where the local energy and net-baryon densities of an ``average event'' are converted into temperature and baryo-chemical potential. They are used as input parameters for a medium modifications e.g. when calculating electromagnetic radiation \cite{endres2016energy,endres2016photon,staudenmaier2018dilepton}. This improves agreement with experimental measurements as shown for dilepton radiation with the SMASH approach in \cite{staudenmaier2018dilepton}. A description of the resonance dynamics solely based on on-shell propagation is not sufficient; instead, a mixed approach including the coarse-graining method leads to a better agreement for large collision systems. 

Nevertheless, even with a resonances description based on vacuum properties, the dynamical evolution of resonances is dramatically different in vacuum and in medium. Of particular interest for this work is the shortening of resonance lifetimes by inelastic scatterings i.e. absorption inside the medium, often referred to as \emph{collisional broadening}. A shortening of the lifetime ($\tau$) translates to an effective, dynamically-generated increase of the width ($\Gamma$) and subsequently broadening of the spectral function ($\Gamma_{\rm eff}={1}/{\tau_{\rm eff}}$).

The goal of this work is to investigate this dynamical broadening on the example of the $\rho$ meson quantitatively with the transport approach SMASH. The effective width and spectral function is reconstructed by analyzing the resonance lifetimes. Both equilibrium and non-equilibrium systems are studied and compared to assess the role of the different dynamics.  

This paper is organized as follows: Sec. \ref{sec:SMASH} describes SMASH, the hadronic transport approach used in this work, and details its relevant features. Sec. \ref{sec:Broadening} defines how the collisional broadening is calculated, with the results for different scenarios presented in Sec. \ref{sec:results} and summarized in Sec. \ref{sec:Conclusion}.

\section{Model Description}\label{sec:SMASH}

For the investigation of the $\rho$ meson in this work, the hadronic transport approach SMASH-2.1 \cite{weil2016particle,dmytro_oliinychenko_2021_5796168} is employed. It allows to access the full phase space information at all times. Particles can be followed individually and their lifetimes and interactions are directly accessible. Hadrons evolve in spacetime according to an effective solution of the relativistic Boltzmann equation. 
Particle species, their pole masses $M_0$, and corresponding decay widths $\Gamma_0$ are taken from the Particle Data Group \cite{Zyla:2020zbs} up to $M_0\sim2.3\ \mathrm{GeV}$. Hadrons with $\Gamma_0\leq10\ \mathrm{keV}$ are considered stable, otherwise they are regarded as resonances with a non-singular vacuum spectral function, and can decay with probability given by the mass-dependent decay width. 
Only two-body decays and scatterings are included in the calculations with a geometrical collision criterion in order to maintain detailed balance. Resonances go through a $1\to2$ decay, or are absorbed, either in a $2\to1$ resonance formation or a $2\to2$ inelastic collision \cite{weil2016particle}.

Equilibrium properties are studied in infinite matter calculations. This is achieved by using a finite box with periodic boundary conditions. The initial multiplicities are sampled from a Poisson distribution, simulating a grand-canonical ensemble. Particle momenta are sampled from a Maxwell-Boltzmann distribution with the given $(T,\mu_B)$, which approximates thermal and chemical equilibrium. 
 For nucleus-nucleus collisions, the nuclei travel towards each other along the longitudinal axis with a given kinetic energy, and offset in the transverse axis by a given impact parameter. For the initial condition, the positions of nucleons in each nucleus are sampled according to the Woods-Saxon distribution, without Fermi momentum.
Densities in this work are computed in the Eckart rest frame with a Gaussian smearing. For the hadron density, each particle in SMASH has the same unitary weight.

\subsection{Decay widths}

The hadronic decay widths in SMASH follow the treatment of \cite{manley1992multichannel}, where the two-body decay $R\to ab$ has a mass-dependent width of
\begin{equation}\label{SMASH:partial_width_def}
\Gamma^\mathrm{dec}_{R\to ab}(m)=\Gamma_{R\to ab}^0\frac{\rho_{ab}(m)}{\rho_{ab}(M_0)},
\end{equation}
where $m$ is the off-shell mass, $M_0$ and $\Gamma_{R\to ab}^0$ are the pole mass and corresponding width, and $\rho(m)$ is a parametrization, described in full detail in \citep{weil2016particle}.

The proper lifetime of a resonance with mass $m$ is defined as 
\begin{equation}\label{SMASH:proper_lifetime_def}
\tau=\frac{1}{\Gamma^\mathrm{dec}(m)},
\end{equation}
where $\Gamma^\mathrm{dec}(m)$ is the \emph{total} decay width, computed as the sum of the partial widths \eqref{SMASH:partial_width_def} over all decay channels $\{ab\}$. The probability for the resonance to decay in a sufficiently small time interval in its rest frame is \begin{equation}\label{SMASH:prob_decay}
P(\mathrm{decay\ in\ }\Delta t)=\frac{\Delta t}{\tau}=\Gamma^\mathrm{dec}(m)\Delta t.
\end{equation}

SMASH uses this probability at each timestep to decay resonances. When it happens, a decay channel $R\to X$ is  randomly chosen from the list of possible processes, with a mass-dependent branching ratio $\Gamma^\mathrm{dec}_{R\to X}(m)/\Gamma^\mathrm{dec}(m)$.

\subsection{Spectral function}

The spectral function of a resonance relates to the imaginary part of its propagator, and hence carries information about the mass probability distribution. In general, it can depend on temperature and density; however, such in-medium modifications are currently neglected in SMASH, and vacuum properties are assumed for all particles. The spectral function of a resonance is given by the relativistic Breit-Wigner distribution
\begin{equation}\label{SMASH:Breit-Wigner}
\mathcal{A}(m)=\frac{2\mathcal{N}}{\pi}\frac{m^2\Gamma^\mathrm{dec}(m)}{(m^2-M_0^2)^2+m^2\Gamma^\mathrm{dec}(m)^2},
\end{equation}
where $\mathcal{N}$ is a normalization factor. 
When a resonance is formed, SMASH samples its mass using \eqref{SMASH:Breit-Wigner} within the available phase-space. The mass distribution in a simple gas in equilibrium, for instance, amounts to folding \eqref{SMASH:Breit-Wigner} with a thermal distribution:
\begin{equation}\label{SMASH:eq_folding}
\frac{1}{N}\dv{N}{m}\ (m;T,\mu)\propto(mT)^{3/2}e^{(\mu-m)/T}\mathcal{A}(m).
\end{equation} 

Depending on kinematic limitations, different channels for the production of a $\rho$ will be available, which can lead to interesting non-thermal structures in the mass distribution \cite{schumacher2006theoretical,vogel2006reconstructing}.

\section{Collisional broadening}\label{sec:Broadening}

In a hadronic medium, absorption of particles decreases the average lifetime of resonances compared to that in vacuum. Such a decrease can be considered as an effective increase of the total decay width, consequently widening the spectral function. This effect is known as \emph{collisional broadening}. In a medium, absorptions are the main mechanism that determines resonance lifetimes. 

The effective total width is computed by extracting the average lifetime of a collection of particles, and inverting \eqref{SMASH:proper_lifetime_def} to define
\begin{equation}\label{Broadening:Gamma_eff_def}
\Gamma^\mathrm{eff}=\frac{1}{\avg{\tau}}=\avg{\frac{t_f-t_i}{\gamma}}^{-1},
\end{equation}
where $\gamma$ is the Lorentz factor of the resonance with respect to the computational system, computed with the momentum of the resonance, which is taken from the interaction history provided by SMASH along with the initial and final times $t_{i,f}$. The average in \eqref{Broadening:Gamma_eff_def} can be computed differentially, for instance depending on the invariant mass or the local hadron density. The additional contribution to the width, the collisional width, is defined as $\Gamma^\mathrm{col}=\Gamma^\mathrm{eff}-\Gamma^\mathrm{dec}$. 

In order to study the dynamical effects on the spectral function, the mass-dependent $\Gamma^\mathrm{eff}(m)$ replaces the regular vacuum decay width $\Gamma^\mathrm{dec}$ in \eqref{SMASH:Breit-Wigner}. As particles have a shorter average lifetime in the medium due to absorption processes, $\Gamma^\mathrm{eff}$ is larger than $\Gamma^\mathrm{dec}$. The obtained ``dynamic'' spectral function $\Adyn(m)$ is therefore broader than ${A}(m)$. 
Since resonances that are annihilated in absorptions do not decay, this quantity can be thought of as the spectral function of $\rho$ mesons which effectively contribute to the dilepton yields in \cite{staudenmaier2018dilepton}. 

In order to allow for a comparison between different systems, the spectral function must be properly normalized. This is not trivial since higher masses are increasingly rare, so the support of $\Adyn(m)$ is not infinite as in the vacuum. Each dynamic spectral function is normalized to the integral of the vacuum Breit-Wigner in the available support. This introduces a small scaling error, which has no impact on the analyses below.

\section{Results}\label{sec:results}

\subsection{Thermal systems}

First, the collisional broadening of $\rho$ mesons is computed inside a box with an equilibrated hadron gas. This allows for an assessment of the thermodynamic behavior of the effective width, as well as for the comparison of $\Adyn$ to well-established model calculations of full in-medium modifications \cite{rapp1999low,van2008dilepton}. The box is set to initialize as explained in Sec. \ref{sec:SMASH} at different temperatures $T\in\{120,150,180\}$ MeV, each with three baryochemical potential values $\mu_B\in\{0,330,450\}$ MeV. Results are only considered after $t=10^4$ fm, which was checked to guarantee  equilibration.

\begin{figure}[ht]
\centering
\includegraphics[width=0.98\linewidth]{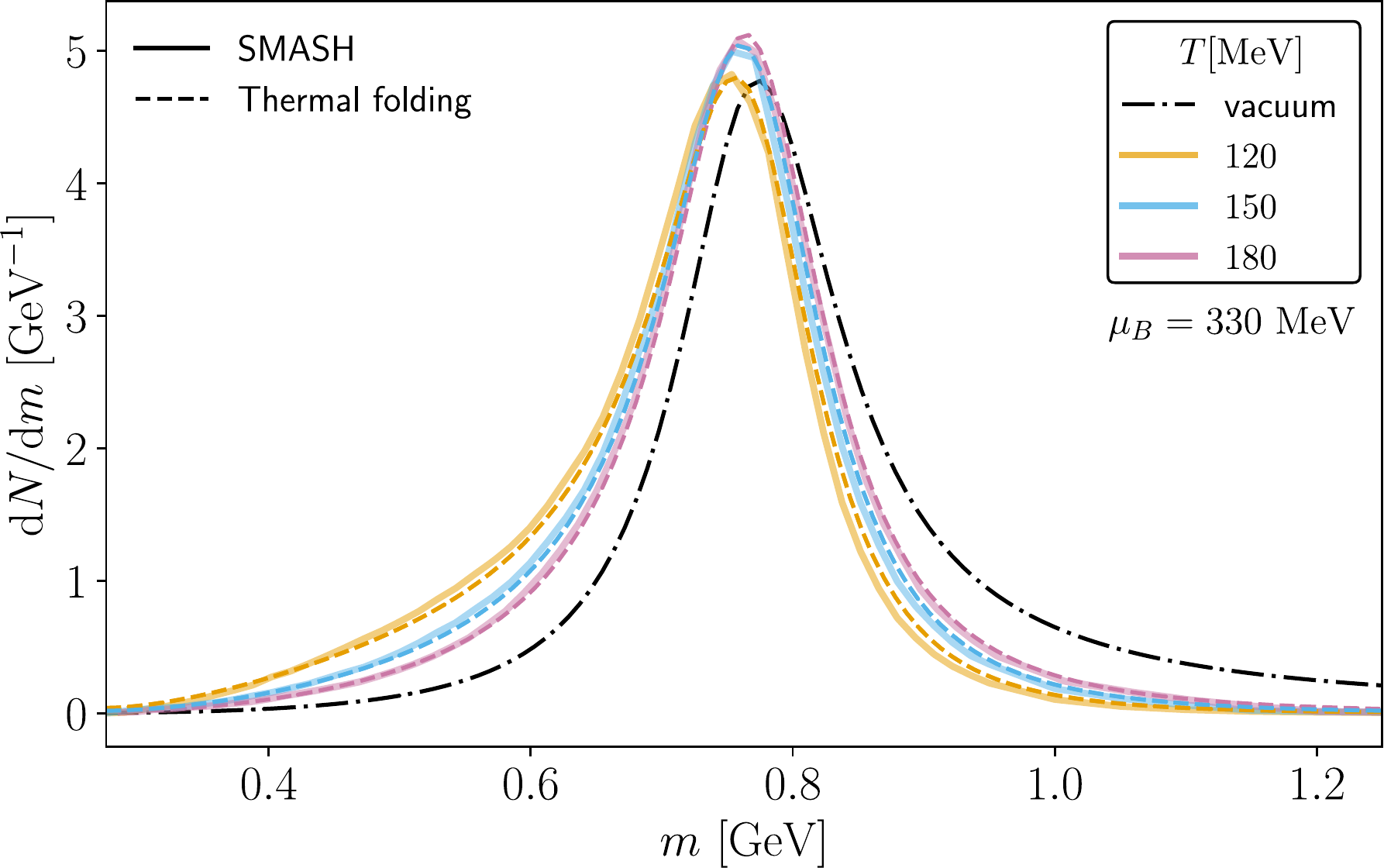}\caption{Probability distribution of $\rho$ meson masses in thermal equilibrium, at $\mu_B=330\ \mathrm{MeV}$ and $T\in\{120,150,180\}\ \mathrm{MeV}.$ SMASH production shown in the solid histogram, and the thermal folding \eqref{SMASH:eq_folding} in dashed lines. The vacuum Breit-Wigner \eqref{SMASH:Breit-Wigner} is shown in black dash-dotted lines.}\label{fig:box_nominal}
\end{figure}

The mass distribution of $\rho$ mesons is shown to closely match the folding \eqref{SMASH:eq_folding} in Fig. \ref{fig:box_nominal}, as the simulated matter indeed corresponds to a thermalized hadron gas. The thermal weight exponentially favors the creation of smaller masses in comparison to the vacuum \eqref{SMASH:Breit-Wigner}, and there is not much difference between the production at the selected temperatures. The value of baryochemical potential does not alter the distribution -- at least in this region of the phase diagram --, therefore only $\mu_B=330$ MeV is displayed. This is consistent with the folding, in which the dependence on $\mu_B$ is cancelled by the normalization. In simpler systems exact matches between the analytic expectation and SMASH were found, and the slight deviations can be attributed to additional production channels in the full hadron gas via the $N^*(1520)$ resonance. 

\begin{figure}[ht]
\centering 
\includegraphics[width=0.92\linewidth]{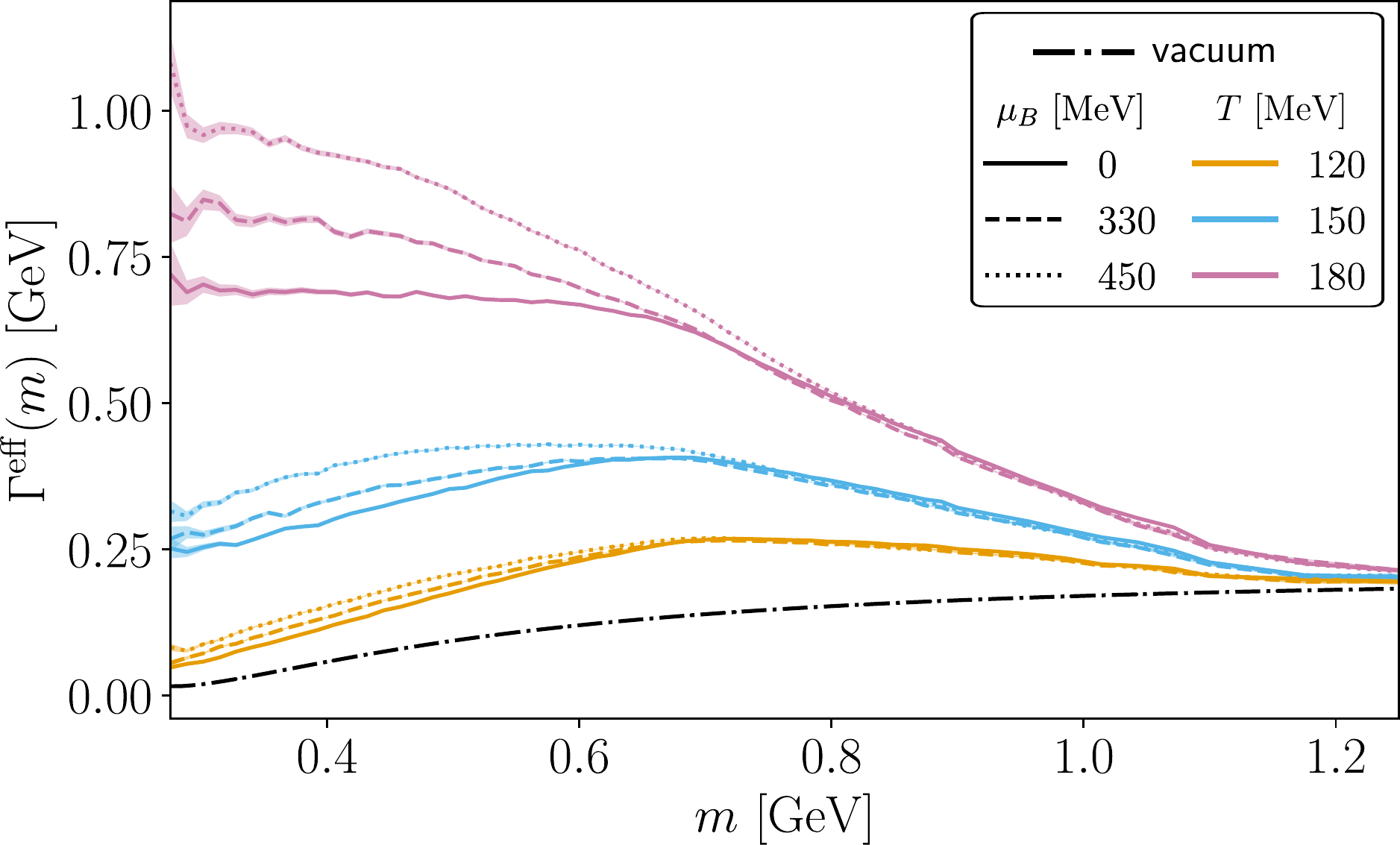}\caption{Effective width of $\rho$ mesons in thermal equilibrium. Error bands are statistical.}\label{fig:box_width}
\end{figure}

Applying \eqref{Broadening:Gamma_eff_def} to the $\rho$ mesons separated in bins of mass, the mass-dependent effective width, shown in Fig. \ref{fig:box_width}, is extracted. 
$\Gamma^\mathrm{eff}(m)$ depends strongly on the thermodynamic conditions of the system, in contrast to the mass distribution of Fig. \ref{fig:box_nominal}. 
Lower masses are more affected by changes in the thermodynamic parameters, since the cross-section for $2\to1$ and $2\to2$ processes decreases with the masses of incoming particles in this energy range \cite{weil2016particle}. 
Hence, these heavier $\rho$ mesons are less likely to be absorbed. The baryochemical potential is only relevant below the pole mass of $M_0=776$ MeV; an increase in $\mu_B$ favors the creation of baryons, suggesting that their coupling to the $\rho$ dominates the low-mass region.

\begin{figure}[ht]
\centering 
\includegraphics[width=0.9\linewidth]{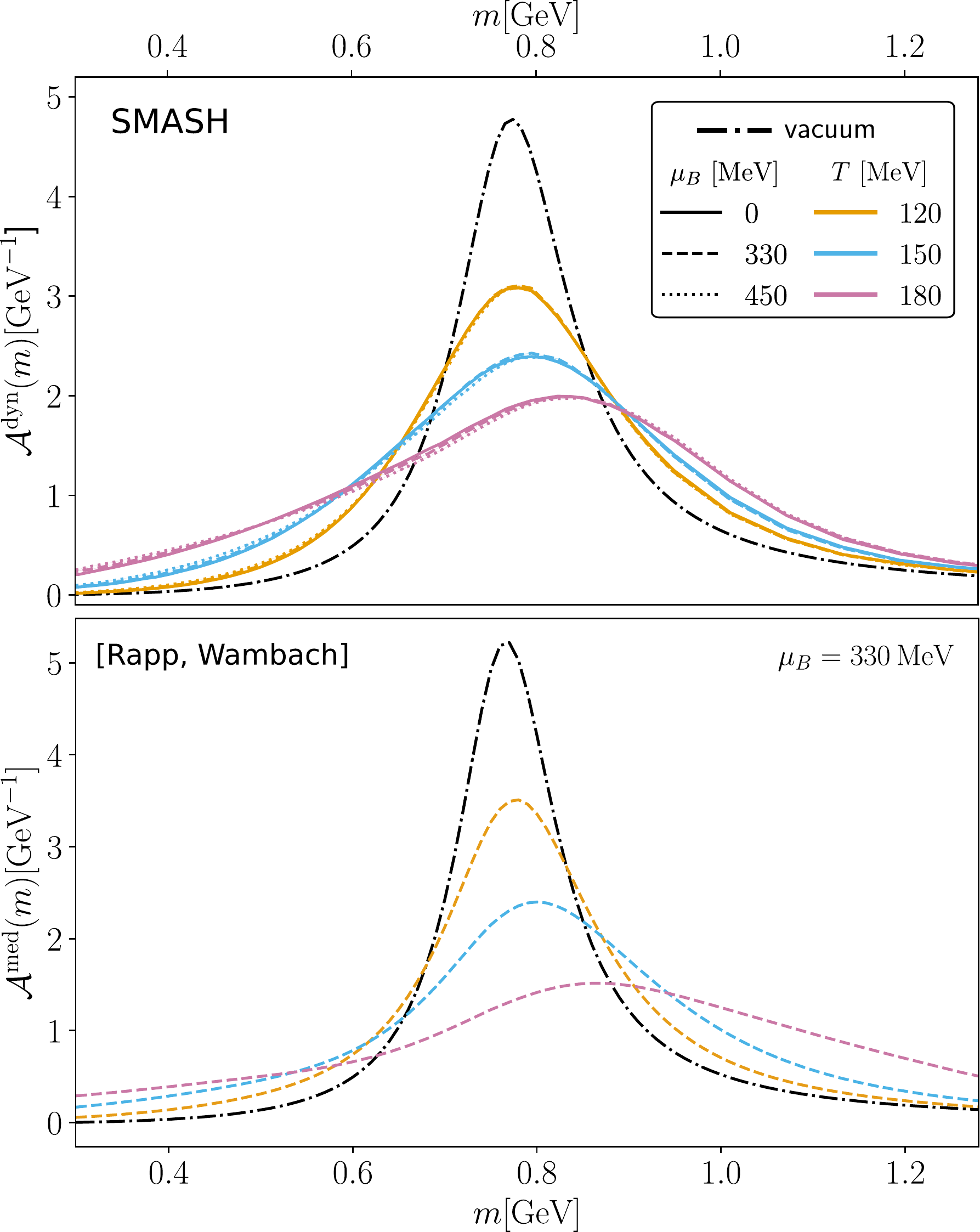}\caption{(Upper) Dynamic spectral function of $\rho$ mesons in thermal equilibrium. (Lower) Momentum-integrated in-medium spectral function for $\mu_B=330$ MeV \cite{rapp1999low}.}\label{fig:box_dynamic}
\end{figure}

Figure \ref{fig:box_dynamic} shows a comparison between the dynamic spectral function from SMASH and the modifications to the $\rho$ meson propagator from a full in-medium model \cite{rapp1999low}. Qualitatively, the behavior of $\Adyn(m)$ is similar to the full in-medium spectral function: it is broadened with increasing temperature, and the peak mass shifts slightly. Notably, the baryochemical potential makes almost no difference in the dynamic spectral function. 

However, there are quantitative differences present. The high-mass tail is less broadened in SMASH, and the opposite happens in the low-mass tail. This is likely due to the ``tree-level'' character of hadronic transport. Quantum corrections -- loops -- are taken into account in the matching of elementary cross-sections to experimental data; that is, only vacuum corrections are correctly described. The medium effects on these diagrams are not present, unlike in the in-medium model description, which modifies the propagator self-consistently including interference terms.

\subsection{Collision systems}

Next, the emergence of collisional broadening is studied in the off-equilibrium matter created by low-energy nuclear collisions following the selection of the HADES collaboration. Experiments assess the medium effects in these systems, by analyzing the excess dilepton yields in comparison to a hadronic cocktail \cite{arnaldi2006first}. The present study is restricted to low beam energies, where the evolution is appropriately described by the kinetic transport approach. Several nuclear systems are considered to obtain insight about how the broadening depends on system size, centrality, and beam energy.

\begin{figure}[ht]
\centering 
\includegraphics[width=0.98\linewidth]{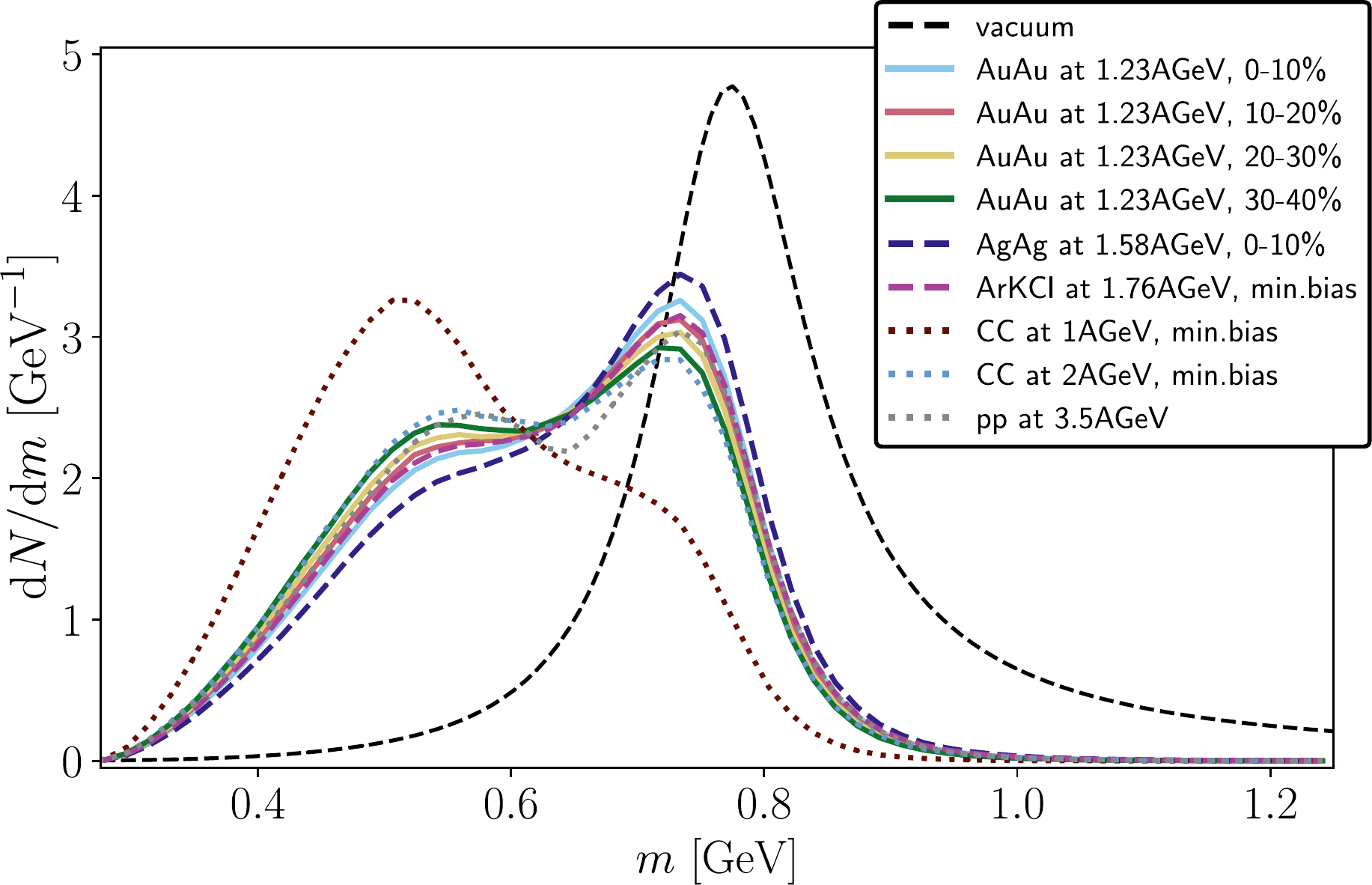}\caption{Nominal spectral function of $\rho$ mesons in different collision systems.}\label{fig:HIC_nominal}
\end{figure}

The mass distribution of $\rho$ mesons, shown in Fig. \ref{fig:HIC_nominal}, for nuclear collisions at HADES beam energies is not following a thermal folding, because the system is very far from equilibrium. There is an enhanced production peak at $\sim0.5\ \mathrm{GeV}$, stemming from the decay $N^*(1520)\to N+\rho $, since the mass of this resonance does not allow to produce a pole-mass $\rho$ \cite{schumacher2006theoretical,vogel2006reconstructing}. This is most apparent in the C+C collisions at $E_\mathrm{kin}=1$ AGeV, where the low beam energy changes the preferred production channel. This effect is purely kinematic, with this peak populated by low-momentum $\rho$ mesons \cite{reichert2022kinetic}, since the production of resonances does not take lifetimes nor in-medium effects into account.

\begin{figure}[ht]
\centering 
\includegraphics[width=0.98\linewidth]{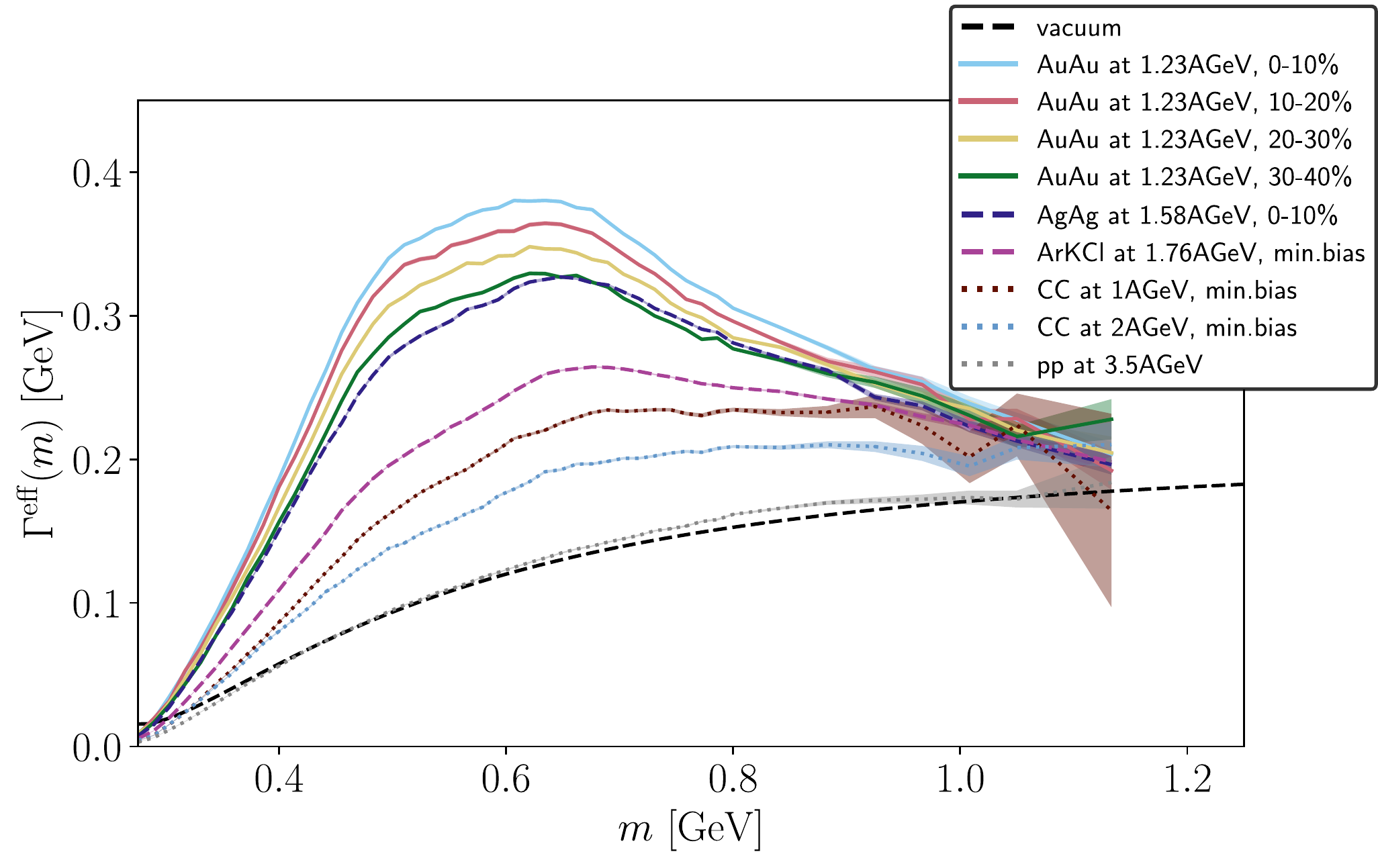}\caption{Effective width of $\rho$ mesons in different collision systems. Error bands are statistical.}\label{fig:HIC_width}
\end{figure}

The effective width shown in Figure \ref{fig:HIC_width} reveals system size as a dominant factor. In more central collision the dynamic width is larger, as evidenced by the four centrality classes of Au+Au collisions. Similarly, heavier nuclei reveal a stronger collisional broadening. The p+p collision follows the vacuum decay, since there is essentially no medium with which the $\rho$ mesons interact formed. 

However, system size is not the only factor, given that a Au+Au collision at 30-40\% broadens as much as a central Ag+Ag, even though it has far fewer participants on average\footnote{Using the Woods-Saxon model with an inelastic NN-cross section of $\sigma_{NN}=30$ mb, a 30-40\% Au+Au collision has $\sim98$ participants on average, whereas a central Ag+Ag collision has $\sim168$.}. The beam energy is playing a role as well: the faster velocity of the ions in the larger system lets the medium dissipate faster, leading to fewer binary collisions overall. This is also seen in C+C collisions, the width at $E_\mathrm{kin}=1$ AGeV is larger than at $2$ AGeV.

\begin{figure}[ht]
\centering 
\includegraphics[width=0.98\linewidth]{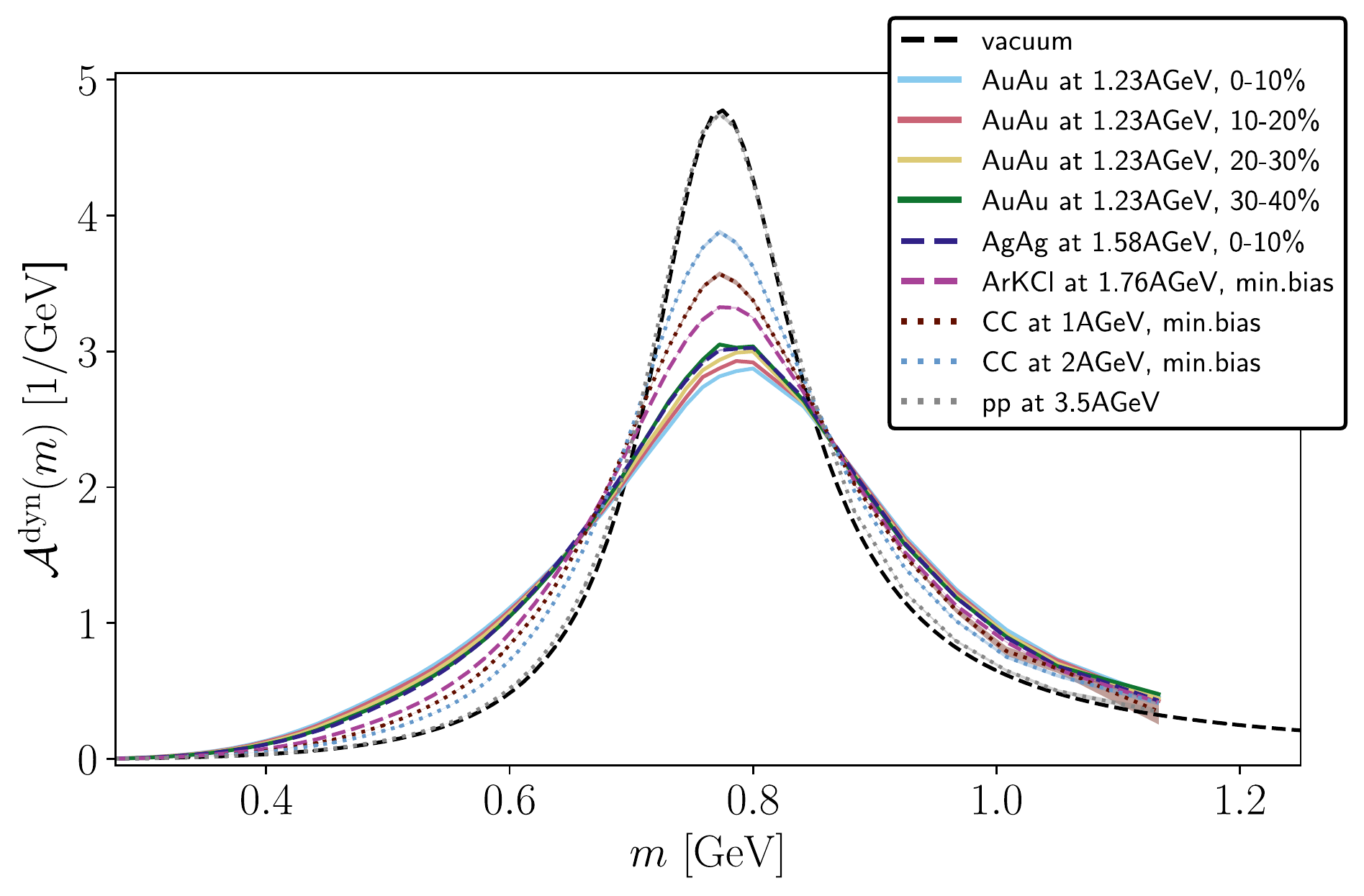}\caption{Dynamic spectral function of $\rho$ mesons in different collision systems.}\label{fig:HIC_dynamic}
\end{figure}

Unlike in the thermal system, the broadening below $m\sim0.5$ GeV always decreases, because most low-mass resonances are created in the later stages of the collision, when the medium is already dilute. Furthermore, the peak mass of the dynamic spectral function, shown in Fig. \ref{fig:HIC_dynamic}, does not shift to a great extent, as these nuclear systems do not reach temperatures comparable to the selected box values of Fig. \ref{fig:box_dynamic}.

\subsubsection*{Time evolution}

\begin{figure}[ht]
\centering 
\includegraphics[width=0.98\linewidth]{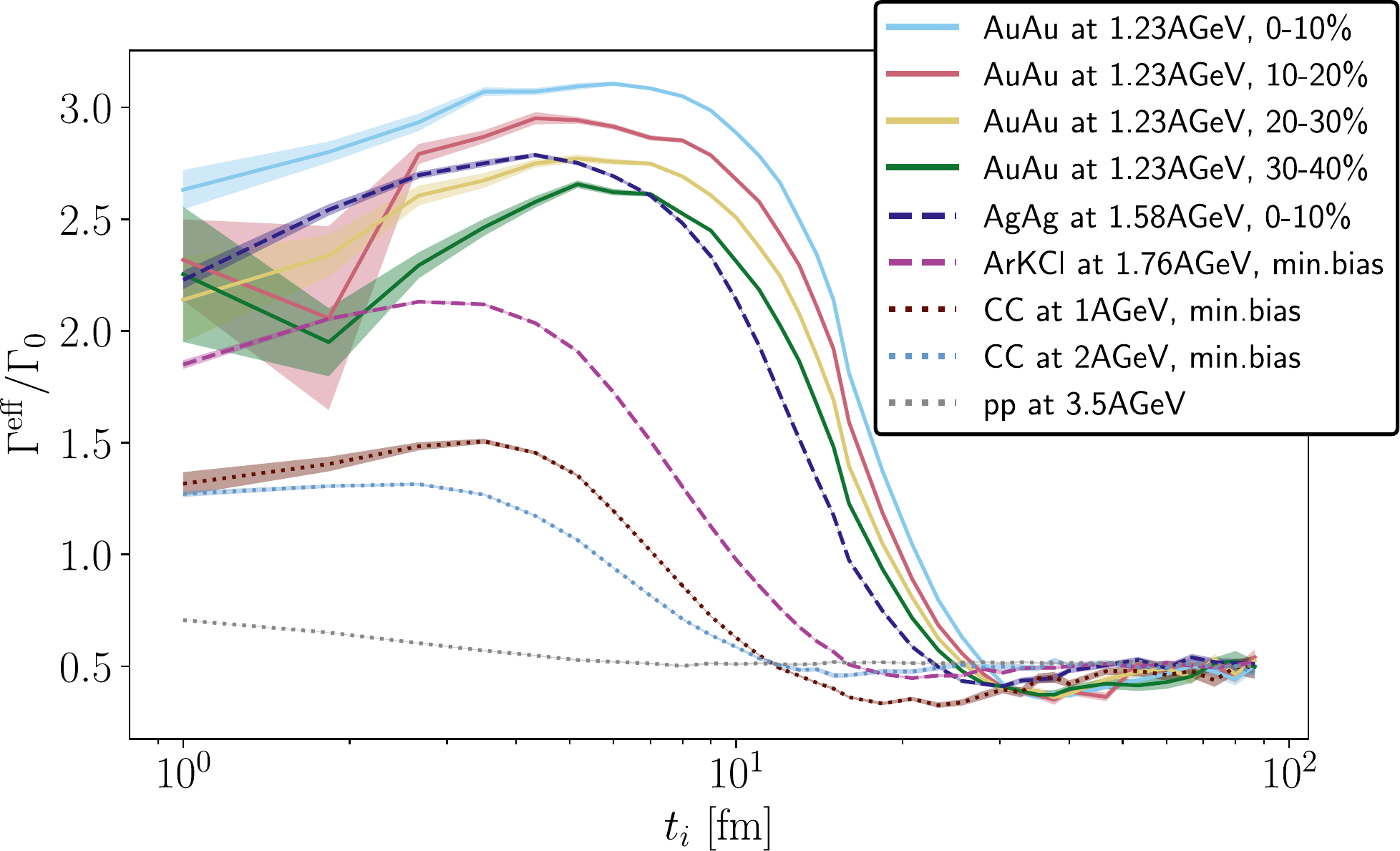}\caption{Time evolution of the effective width of $\rho$ mesons in different collision systems.}\label{fig:HIC_evolution}
\end{figure}

It is also interesting to observe how the collisional broadening evolves with the expansion of the fireball. This is seen by computing \eqref{Broadening:Gamma_eff_def} as a function of the time $t_i$ at which the resonance is created (in the computational frame). The corresponding width is shown in Fig. \ref{fig:HIC_evolution}, normalized by the width at pole.

\begin{figure}[ht]
\centering 
\includegraphics[width=0.98\linewidth]{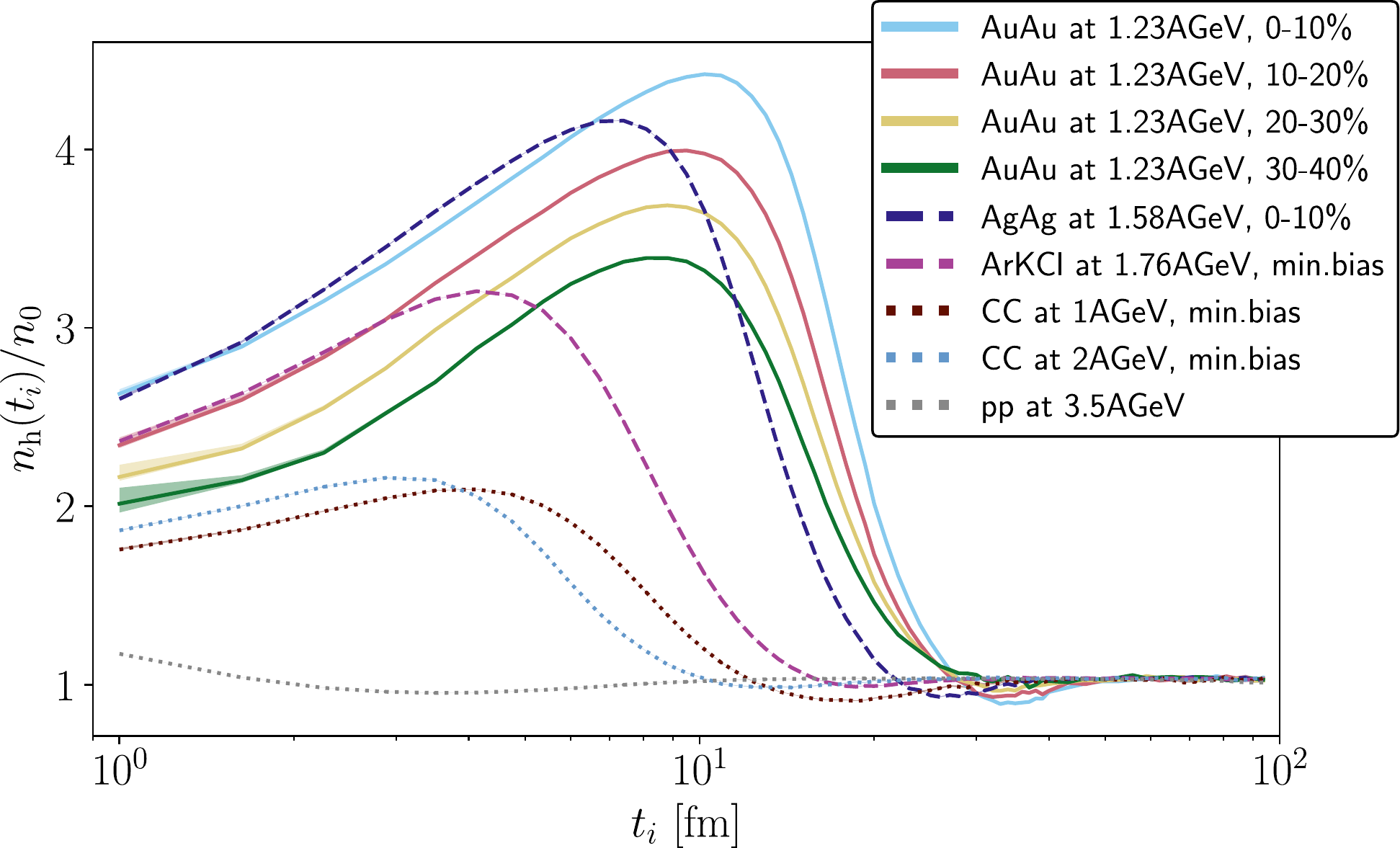}\caption{Time evolution of the average hadron density at the interaction in which a $\rho$ meson is removed from the system.}\label{fig:HIC_evolution_density}
\end{figure}

One way to understand this evolution is by probing the lifetime of the medium itself. This is done in Fig. \ref{fig:HIC_evolution_density}, where $n_0=0.16~\mathrm{fm}^{-3}$ is the nuclear ground state density, by using the average hadron density $n_\mathrm{h}$ as a proxy. 
The densities in SMASH are computed at interactions, so the hadron density is chosen to be calculated at $t_i$.

During the first few fm of the collision, the effective width grows as the nuclei traverse each other, until the number of binary collisions is maximal. This occurs when $n_\mathrm{h}$ is largest and so the position of this maximum changes with nuclear mass and beam energy. 

As the medium expands, it becomes more dilute and consequently the effective width decreases. In light of this, the difference in the effective widths of 30-40\% Au+Au and central Ag+Ag systems -- that have a similar mass-dependent width in Figure \ref{fig:HIC_width} -- becomes evident. Having more participating nucleons, the initial broadening in the latter is larger, but it decreases faster than in the former.

\begin{figure}[ht]
\centering 
\includegraphics[width=0.98\linewidth]{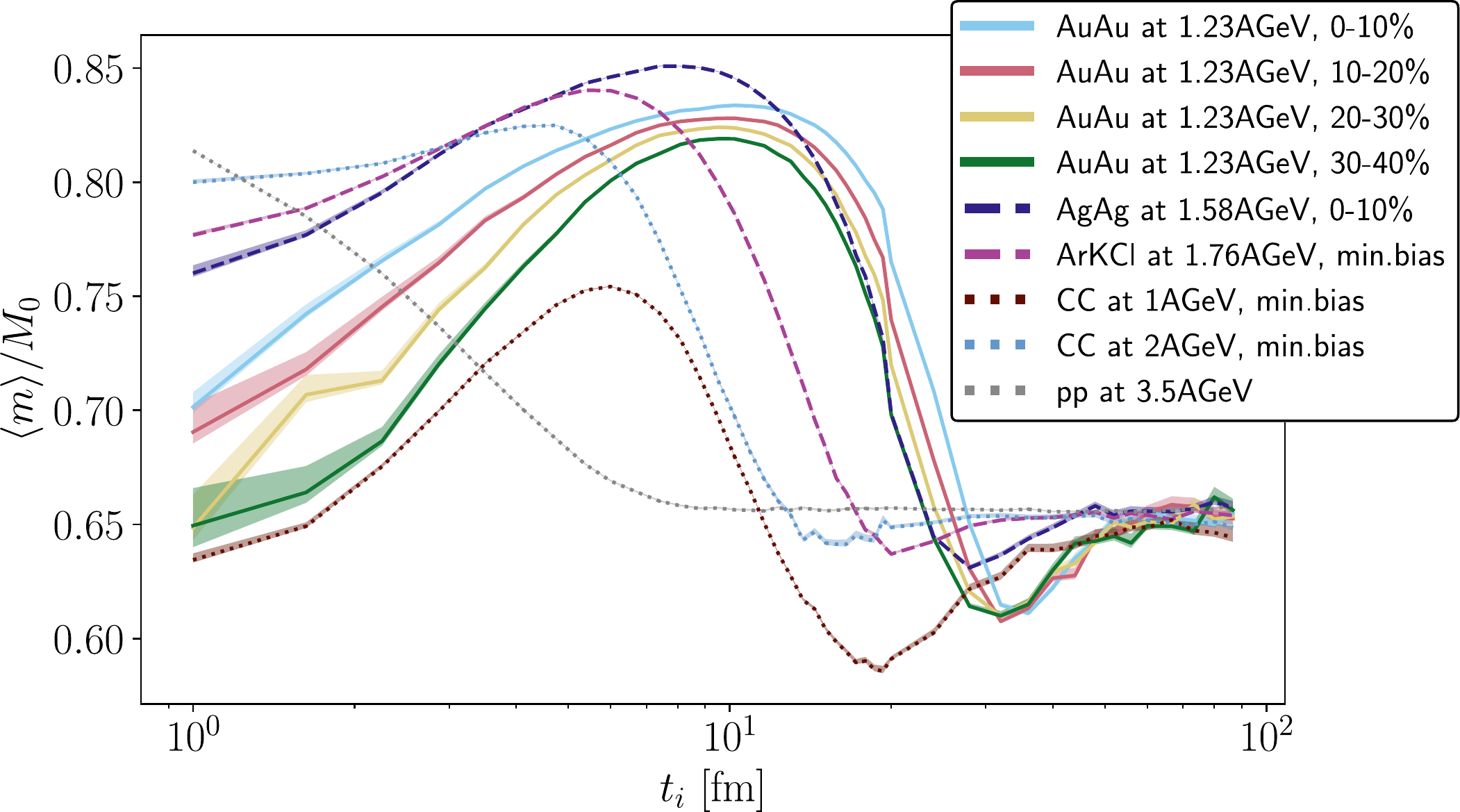}\caption{Time evolution of the average $\rho$ meson mass in different nuclear collisions.}\label{fig:HIC_evolution_mass}
\end{figure}

At later times, particles are essentially travelling in vacuum, hence $\Gamma^\mathrm{eff}\approx\Gamma^\mathrm{dec}$. Since higher masses have higher vacuum decay widths (as per Figs. \ref{fig:box_width} and \ref{fig:HIC_width}), and  $\avg{m}<M_0$ as seen in Fig. \ref{fig:HIC_evolution_mass}, the widths in all systems fall to small values less than $\Gamma_0=0.149$ GeV. This also explains why, for instance, the width in a C+C collision falls below the one in p+p at this stage.

\subsubsection*{Density dependence}

\begin{figure}[ht]
\centering 
\includegraphics[width=0.98\linewidth]{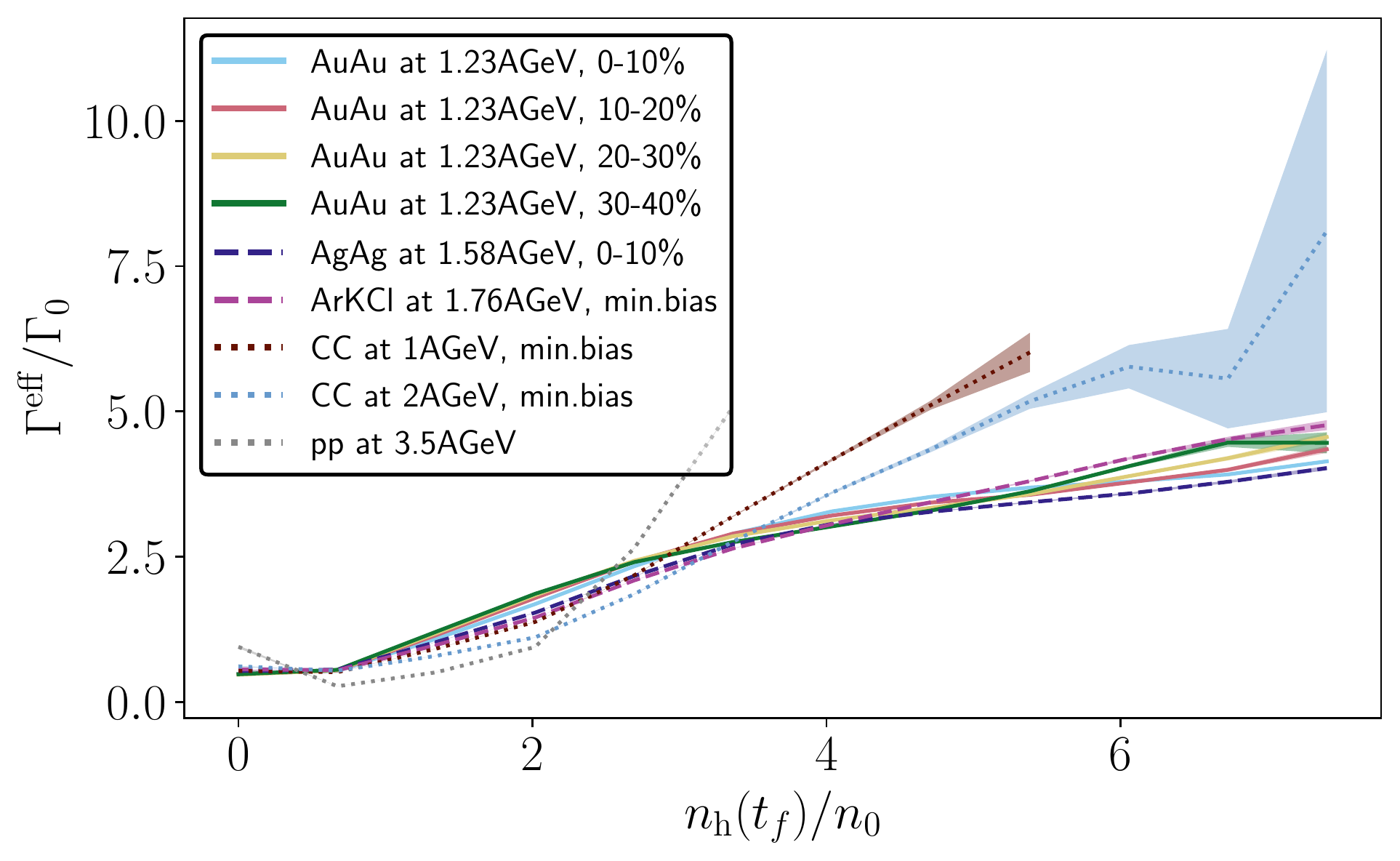}\caption{Density dependence of the effective width of $\rho$ mesons in different collision systems.}\label{fig:HIC_width_density}
\end{figure}

Figures \ref{fig:HIC_evolution} and \ref{fig:HIC_evolution_density} in the previous subsection suggest a monotonic dependence of the effective width on the local hadron density, as seen in Figure \ref{fig:HIC_width_density}. Here, the density is computed at the end-interaction (at time $t_f$) of the resonance, so as to probe the medium conditions where the $\rho$ is absorbed or decays. This is an approximate way of probing $\Gamma^\mathrm{eff}(n_\mathrm{h})$, as it does not consider how the density changes during propagation.

A near universal relation is recognizable, with most systems following similar curves, while in Figs. \ref{fig:HIC_width} and \ref{fig:HIC_evolution} the widths of the $\rho$ mesons are distinct in different systems. This universality is reminiscent of the semiclassical calculation $\Gamma^\mathrm{col}\sim \gamma n_N\avg{v\sigma_{VN}^\mathrm{tot}}$, used in models in which the mass is not constant and the collisional width is proportional to the local nucleon density \cite{bratkovskaya2008dilepton, larionov2020dilepton}. 

A deviation to this appears in the p+p and C+C  systems, when the density is high. This is because the density calculation starts to break down for very dilute system and an influence of the reaction partners in the specific binary interaction cannot be ruled out as there are never many particles close to the interaction point, potentially being the reason for the spurious high densities seen for such small systems.

\subsection{Non-equilibrium effects}

Having quantified the collisional broadening in these two different scenarios, it is tempting to ask: how does the effective width in a non-equilibrated collision system compare to the thermalized value of a box? 

\begin{figure}[ht]
\centering
\includegraphics[width=0.95\linewidth]{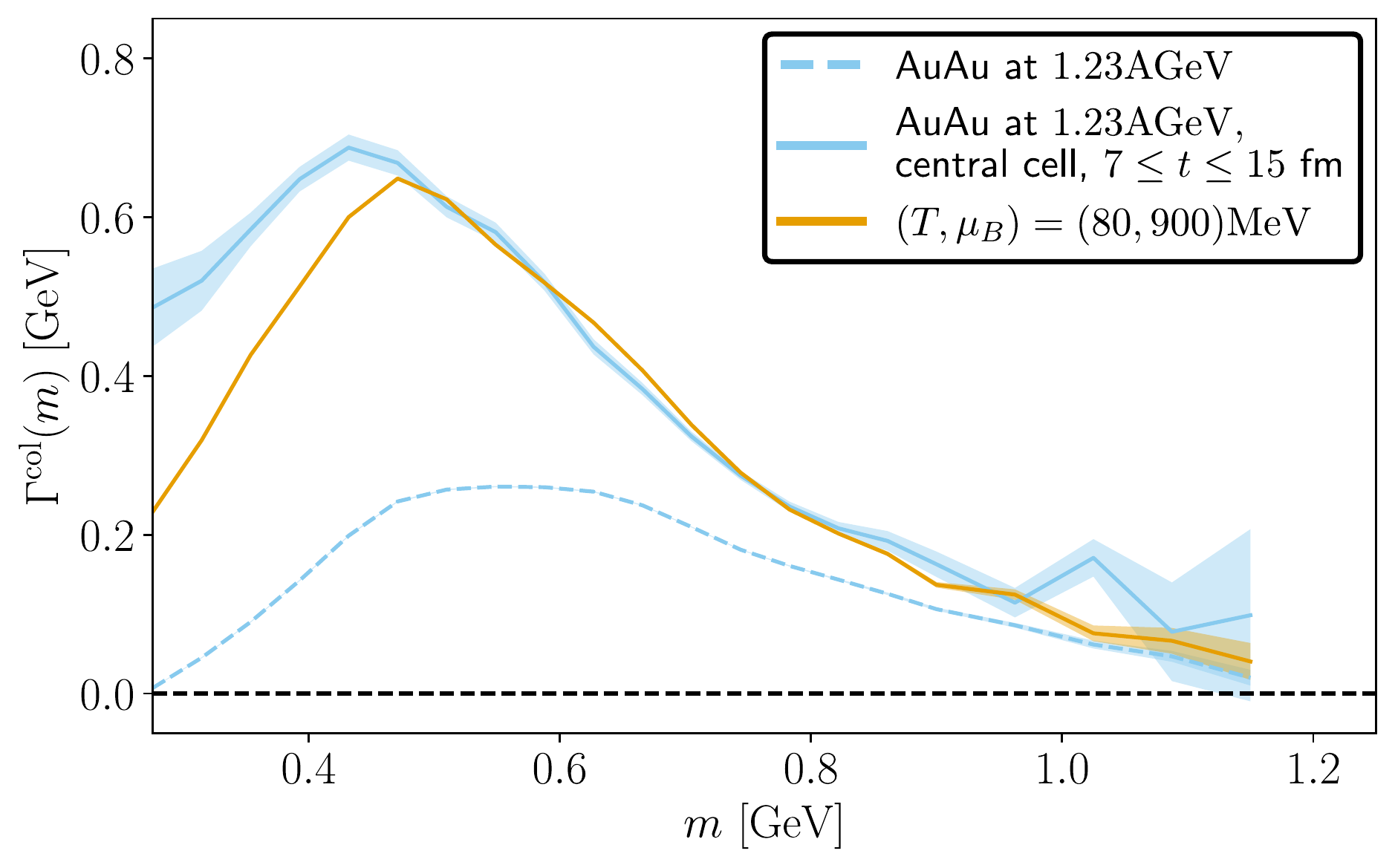}\caption{Collisional width of a central Au+Au collision at $E_\mathrm{kin}=1.23$ GeV, in the full phase and restricted to a region of constant $(T,\mu_B)$; in a box initialized with these thermodynamic parameters.}\label{fig:noneq_width}
\end{figure}

There is a spacetime region of a heavy-ion collision in which the temperature and chemical potential\footnote{These are computed by assuming the HRG equation of state with local densities of energy and baryon number.} are nearly constant \cite{staudenmaier2018dilepton}. This allows to compare the effective width in a collision system with a box at the same thermodynamic conditions (shown in Figure \ref{fig:noneq_width}). In the central cell of a Au+Au collision, the system remains stable in the same thermodynamic state of around $(T,\mu_B)=(80,900) \mathrm{MeV}$ early in the collision, between $7\leq t\leq15 \ \mathrm{fm}$ \cite{staudenmaier2018dilepton}. The system is much denser on average in this spacetime region, so the width is correspondingly enhanced. Since this restriction removes the late-stage resonances, the collisional width at the hadronic threshold is no longer approximately zero. A thermalized hadron gas at the same $(T,\mu_B)$ manifests a similar broadening for masses above $0.5$ GeV. Below this value, the collision system displays a larger broadening, meaning that the non-equilibrium character in a heavy-ion system results in a suppresion of lifetimes of low-mass resonances.

The suppression is explained by considering the hadronic content of each scenario in Fig. \ref{fig:noneq_species}. The invariant yield is computed by considering the average number of the particles over the relevant region of the phase space, and normalizing by the size of that region. The species displayed are first ordered by multiplicity in the central cell of the Au+Au collision until $N^*(1535)$, then by the multiplicity of remaining species in the box.

During the interval of constant thermodynamic conditions, the central cell of the nuclear system is composed mostly of baryons, being dominated by nuclear and $\Delta$-baryon resonances. As discussed in the introduction  \cite{agakichiev2010origin,rapp2009chiral,salabura2021dilepton}, these couple strongly with low-mass $\rho$ mesons, enhancing the effective width. These particles are not as abundant in the thermal gas. Instead, the energy is distributed in the form of lighter hyperons, since the multiplicity is $\propto e^{-m/T}$ in an equilibrated system. Such particles do not couple as much with the $\rho$, and hence do not increase the width in the thermal box.

\section{Conclusion}\label{sec:Conclusion}

In this work, the collisional broadening of $\rho$ mesons was investigated and quantified by computing their effective width via lifetime analysis. The employed transport approach, SMASH \cite{weil2016particle}, relies on vacuum properties of hadrons, so that the mass distribution is given by a vacuum Breit-Wigner function adjusted to the kinematically available energy. The ``dynamic'' spectral function is computed using the effective width, as the collisional broadening emerges from absorption processes as part of the evolution of the hadronic medium. 

First, an infinite hadron gas in equilibrium has been calculated for different temperatures and baryochemical potentials. The resulting spectral functions is compared to full in-medium model calculations \cite{rapp1999low}, which take into account modifications to self-energies and higher order interactions and a qualitatively similar broadening is observed, suggesting that a significant contribution to medium modifications of the $\rho$ meson comes from hadronic interactions irrespective of chiral symmetry restoration, which is not directly present in SMASH. Numerically, however, the shapes in Fig. \ref{fig:box_dynamic} are different. This evidences that the collisional broadening alone is not sufficient to reproduce the experimental dilepton yield in heavy-ion collisions, so other methods using full in-medium modifications need to be applied, such as coarse-graining \cite{staudenmaier2018dilepton}.

Furthermore, the emergence of collisional broadening is studied in non-equilibrium systems created by nuclear collisions (pp, CC, ArKCl, AgAg, AuAu) at different beam energies and centrality classes. The effective width exhibits a clear dependence on system size, as a larger medium enhances the broadening. It also reveals that larger beam energies leads to smaller widths, which is understood through the time evolution of the system. These observations are caused by an universal dependence on the local hadron density. 

Lastly, the two scenarios are compared in order to assess non-equilibrium effects. This has been achieved by simulating a box with similar thermodynamic conditions to those present in a spacetime region of Au+Au collisions. Above $m\sim0.5$ GeV the observed broadening is similar, while below it the collision system exhibited an enhanced width, which is explained by the different hadronic composition of the two scenarios.

The lifetime analysis used throughout this work can be used to understand how inelastic scatterings of vector mesons affect the decay width dynamically and quantify this effect in a transport approach in contrast to genuine in-medium modifications. In the future, it will be interesting to investigate if assumptions about the resonance properties, such as the exact way to calculate the decay probability also affect the effective width and therefore potentially the emission of dileptons. 

\section*{Acknowledgments}

This work was supported by the Helmholtz Forschungsakademie Hessen für FAIR (HFHF) and in part by the National Science Foundation (NSF) within the framework of the JETSCAPE collaboration, under grant numbers ACI-1550228 and OAC-2004571. The authors also acknowledge the support by the State of Hesse within the Research Cluster ELEMENTS (Project ID 500/10.006), and by the Deutsche Forschungsgemeinschaft (DFG, German Research Foundation) – Project number 315477589 – TRR 211. Computational resources have been provided by the GreenCube at GSI. 

\onecolumngrid\
\begin{center}\
\begin{figure}[h]\
            \includegraphics[width=0.75\linewidth]{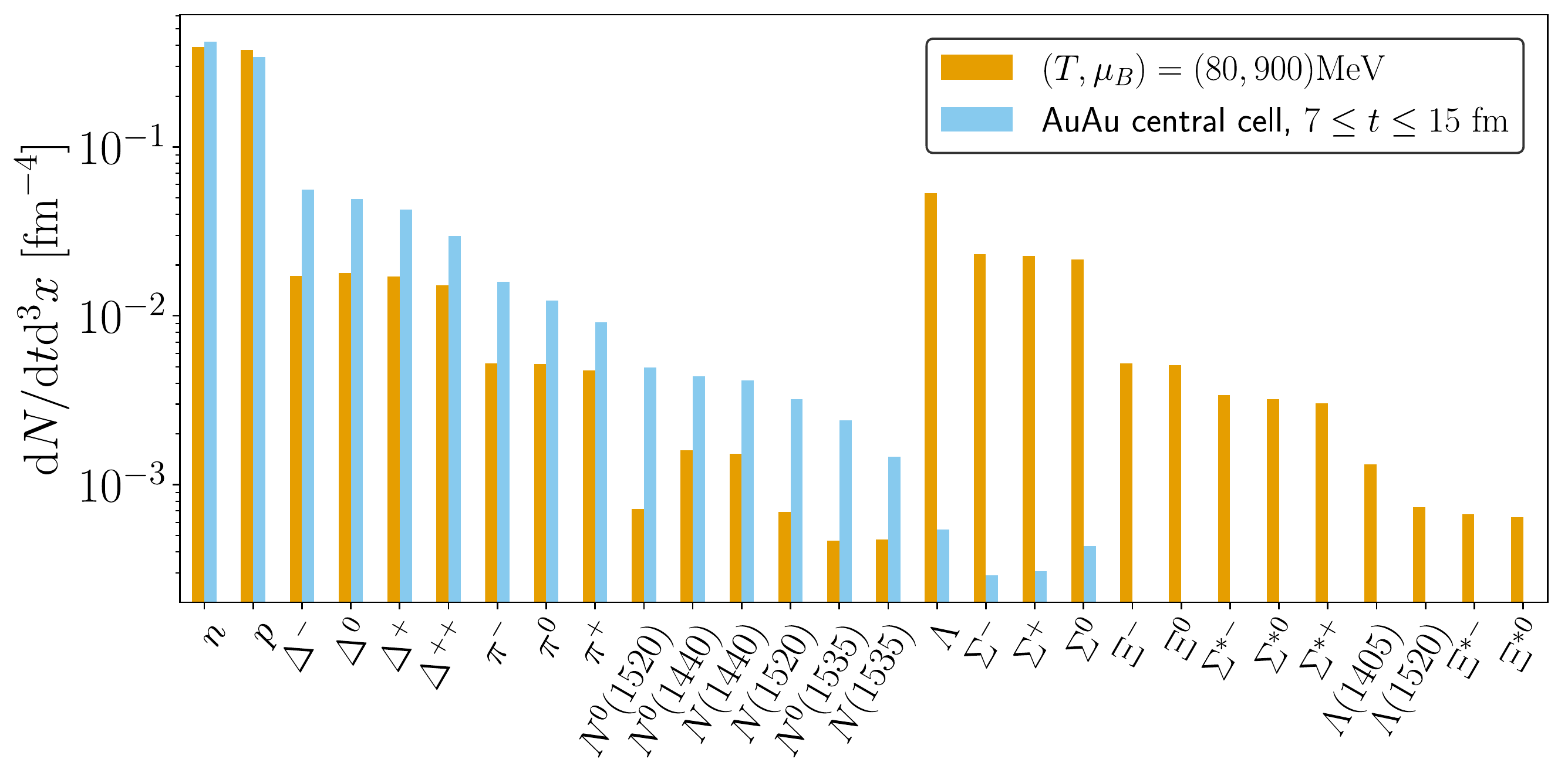}\caption{Multiplicity of each species in the thermodynamically stable region of a Au+Au collision, and in a box under the same $(T,\mu_B)$.}
            \label{fig:noneq_species}
\end{figure}\
\end{center}\
\twocolumngrid

\bibliography{mybib}

\end{document}